# Two-dimensional silk

Chenyang Shi[1,2], Marlo Zorman[3], Xiao Zhao[4,5], Miquel B. Salmeron[4], Jim Pfaendtner[6], Xiang Yang Liu[7]*, Shuai Zhang[1,2]*, James De Yoreo[1,2]*

[1]Physical Sciences Division, Pacific Northwest National Laboratory; Richland, WA, USA.

[2]Department of Materials Science and Engineering, University of Washington; Seattle, WA, USA.

[3]Molecular Engineering & Science Institute, University of Washington; Seattle, WA, USA.

[4]Materials Sciences Division, Lawrence Berkeley National Laboratory; Berkeley, CA, USA

[5]Department of Materials Science and Engineering, Stanford University; Stanford, CA, USA
[6]Department of Chemical & Biomolecular Engineering, North Carolina State University; Raleigh, NC, USA

[7]College of Ocean and Earth Sciences, Xiamen University; Xiamen, Fujian, China

*Corresponding author. Email: james.deyoreo@pnnl.gov (J.J.D.Y.); zhangs71@uw.edu (S.Z.); liuxy@xmu.edu.cn (X.L.).

**Abstract:** The ability to form silk films on semiconductors, metals, and oxides or as free-standing membranes has motivated research into silk-based electronic, optical, and biomedical devices[1]. However, the inherent disorder of native silk limits device performance[2,3]. Here we report the creation of highly ordered two-dimensional (2D) silk fibroin (SF) layers on van der Waals solids. Using in situ atomic force microscopy, synchrotron-based infrared spectroscopy, and molecular dynamics simulations, we develop a mechanistic understanding of the assembly process. We show that the films consist of lamellae having an epitaxial relationship with the underlying lattice and that the SF molecules exhibit the same β-sheet secondary structure seen in the crystallites of the native form. By increasing the SF concentration, multilayer films form via layer-by-layer growth, either along a classical pathway in which SF molecules assemble directly into the lamellae or, at sufficiently high concentrations, along a two-step pathway beginning with formation of a disordered monolayer that subsequently converts into the crystalline phase. Kelvin probe measurements show that these 2D SF layers substantially alter the surface potential. Moreover, the ability to assemble 2D silk on both graphite and $MoS_2$ suggests that it may provide a general platform for silk-based electronics on vdW solids[4].

**Main**

Silk is a natural protein-based material that has been used by humankind for over 5000 years[5]. One of the two main components, silk fibroin (SF) has been exploited in recent decades for its ability to self-assemble into a range of fiber-based architectures that exhibit exceptional mechanical, optical, as well as excellent biocompatibility and biodegradability[1,6,7]. Potential bio-electronic applications have been explored in which SF films are interfaced with vdW solids, metals, or oxides, offering impressive electronic performance for next-generation thin-film transistors, memristors, human-machine interfaces, and sensors[1,8-10], provided that strategies can be developed to address challenges associated with the mismatch between the soft SF proteins and the hard, planar substrates[9].

Previous research to understand the formation and function of SF films has identified the molecular architecture achieved via self-assembly as an essential factor in their properties[1,8,11,12]. Generally, synthesis is carried out through self-assembly in solution followed by deposition on a substrate. However, despite the fact that the interactions between proteins and surfaces, which are absent in bulk solution, have been shown to create unique highly-ordered 2D phases[13-16], this phenomenon remains unexplored in the case of SF films.

Here we report the discovery of a new 2D crystalline phase of SF self-assembled on vdW solids with an epitaxial relationship to the underlying lattice. Using *in situ* AFM (Fig. 1a), synchrotron-based nano-FTIR, and MD, we find that this structure is formed through surface-directed assembly and folding of the SF molecules, with a final architecture comprised of fully ordered monolayers of β-sheet lamellae distinct from the disordered 3D network of β-sheet nano-crystallites that defines silk fibers formed in bulk. Moreover, we show that assembly proceeds along two distinct pathways: direct epitaxial growth at low SF concentrations and a two-step process passing through a transient disordered phase at high concentration. These findings provide a mechanistic understanding of assembly for this canonical biomaterial that can enable efficient approaches to designing and fabricating highly-oriented silk-based bio-electronics.

**Crystalline silk monolayers through heteroepitaxy**

Aqueous SF solutions were made by the dissolution in deionized water of lyophilized SF powder, prepared via the well-established LiBr extraction method[17]. *In-situ* circular dichroism (CD) spectroscopy demonstrates that the SF primary structure in solution immediately after preparation is unstructured with a high content of random coils and negligible β-sheet conformation (Extended Data Fig. 1)[18]. Unlike the 3D fibrils formed when SF solutions are allowed to age, when incubated with HOPG, these SF solutions form well-ordered lamellae (Fig. 1b) with a periodicity of about 5 nm (Fig. 1b and Extended Data Fig. 2a-e) and a height of 0.42 nm (Fig. 1c and Extended Data Fig. 2f-i), which is comparable to the thickness of a single SF fiber having β-sheet secondary structure[19]. AFM imaging also shows that the individual rows are composed of distinct segments (Fig. 1d) and lie along the three armchair lattice directions of HOPG (Fig. 1e-g, and Extended Data Fig. 2j, k). This orientational preference was also predicted by MD simulations in which the SF β-sheet bounding to HOPG was found to be energetically favorable along these directions (Fig. 1h, i and Extended Data Fig. 3). These observations demonstrate that the SF exhibits 2D lattice-matched, epitaxial growth on HOPG.

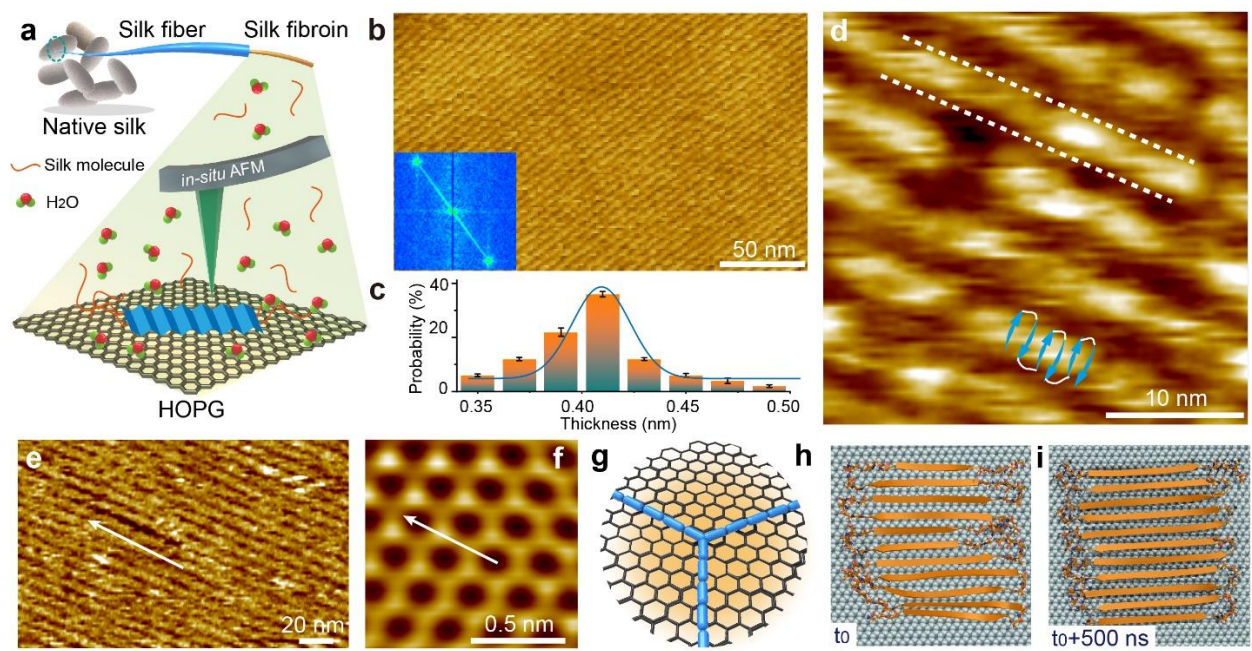

**Fig. 1 | The structure of 2D SF lamellae grown epitaxially on HOPG. a**, Scheme of SF assembly on HOPG characterized by *in-situ* AFM. **b**, AFM image of a lamellar SF monolayer on HOPG formed at an SF concentration of 0.05 μg/ml. Inset shows corresponding FFT pattern. **c**, Thickness distribution of the lamellae. **d**, AFM image of SF lamellae showing distinct segments. The solid blue arrows and white connectors illustrate the proposed array of individual SF β-strands based on MD simulations. Dashed white lines delineate the boundary of a single lamellae. **e-g**, AFM images of (**e**) SF lamellar structure and (**f**) underlying HOPG lattice showing that lamellae lie along the armchair lattice direction as illustrated in (**g**). **h, i**, Simulations of SF β-sheet binding to HOPG. An SF protein with β-sheet conformation initially aligned along the armchair direction (**h**) remains structured in that alignment after 500 ns (**i**).

**Silk multilayers through homoepitaxy**

With increased SF concentration (0.1 mg/ml), multiple lamellar layers formed, with the upper layers again aligned along the armchair directions of the HOPG lattice (Fig. 2a, regions *i* and *ii*). These multilayers exhibited two types of stacking (Fig. 2a, b). In the first, lamellae of one layer are both co-aligned with (Fig. 2b region *i*) and in precise registry with (Extended Data Fig. 4) the lamellae of the underlying layer. As with the first monolayer, these co-aligned lamellae have a height of ~ 0.42 nm (Extended Data Fig. 4c). In the second form of stacking, the upper and lower lamellae are crossed at an angle of 120°, although the height of the lamellae remains the same within error (inset in Fig. 2b).

In conventional silk crystallization from bulk solution, β-sheet strands pack layer-by-layer to form nano-crystallites of coaligned strands that are interconnected within a disordered 3D network[3,20]. Hence, the co-aligned multilayers mimic the bulk crystallization pathway of the nano-crystallites. Moreover, within a given layer, the crossed lamellae are unstable relative to the co-aligned lamellae and, thus, are eventually replaced by co-aligned lamellae through the retreat of the former and the advance of the latter (Fig. 2c). This observation demonstrates that SF monomers are continually undergoing attachment and detachment at the ends of the lamellae and implies there is some degree of misalignment between the amino acid residues of the upper and lower crossed lamellae that interferes with the hydrogen-bond network or the hydrophobic interactions between SF molecules. The resulting film comprises a highly organized sequence of lamellar layers in which β-strands are coaligned within each layer and exhibit precise out-of-plane registry throughout the sequence.

To understand the molecular interactions between the lamellae of adjacent layers, we simulated two possible cases, one in which the β-sheets stack face-to-face — i.e., the well accepted "polar packing" of bulk silk[21] and one in which they stack front-to-back (Fig. 2d, e). The MD simulations predict that both configurations are stable; however, MD simulations of a bilayer with polar packing predict that the height of the second layer is 0.4 nm, similar to what is observed experimentally, while a bilayer with the β-sheets stacked front-to-back is predicted to be more tightly packed and exhibit a height for the second layer of only 0.32 nm (Supplementary Table 1)). These results imply that the observed lamellar films exhibit the same polar packing believed to define the structure of bulk silk fibers.

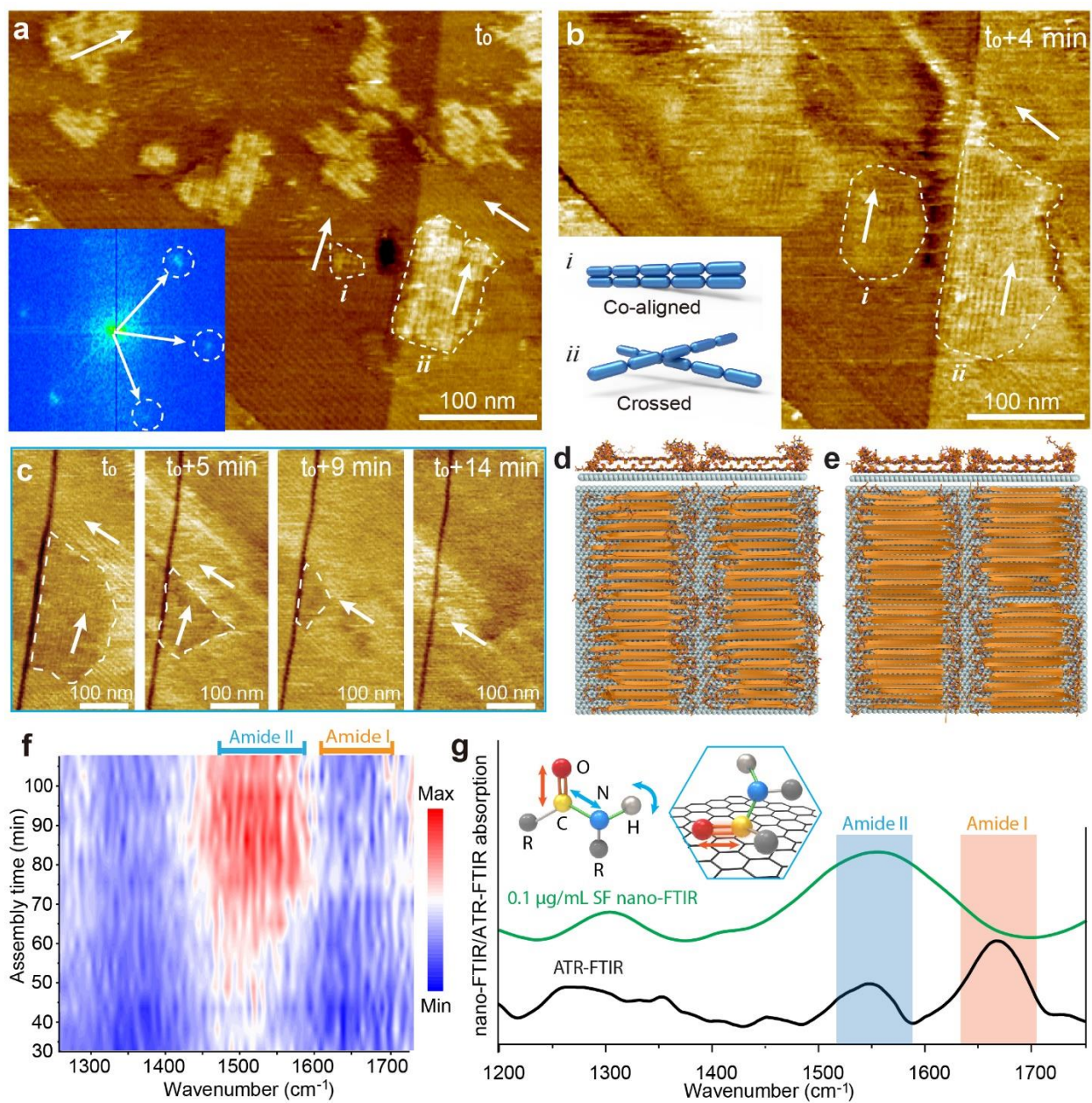

**Fig. 2 | The structure of SF multilayers. a, b,** Two consecutive AFM images of multilayer SF lamellae. The Inset in (**a**) shows the corresponding FFT pattern. White arrows indicate the three directions of the lamellae. The white dashed boundaries delineate two domains (*i* and *ii*) of lamellae overlaying the lower layer. The inset in panel (**b**) illustrates the two possible stacking configurations, co-aligned and at 120° angles. **c,** *In situ* AFM phase images of lamellae, showing the replacement of a crossed domain by a co-aligned domain. The white arrows indicate the orientation within and outside of the crossed domain, which is delineated by the white dashed boundary. **d, e,** MD simulations of bilayer SF lamellae with (**d**) face-to-face and (**e**) front-to-back packing. **f,** Color map showing the time evolution of the *in-situ* nano-FTIR spectra during SF assembly on graphene from an 0.1 μg/ml aqueous SF solution. **g,** *In-situ* nano-FTIR spectra of SF lamellae and *ex-situ* attenuated total reflection Fourier transform infrared (ATR-FTIR) spectrum of dehydrated SF bulk material (See method section for details). The insets illustrate the amide I and II vibration modes during infrared absorption (left) and the expected vibration mode of the ordered amide group on graphene (right), both from the MD simulations of the SF configuration within the lamellae and the lack of an Amide I signal in the nano-IR spectra.

To understand the conformational evolution of the SF molecules during formation of the lamellae, we applied synchrotron-based *in situ* nano-FTIR to follow the change in secondary structure during assembly on graphene with nm resolution (See Extended Data Fig. 5a-c for schematic of the nano-FTIR set-up and example of resulting data). Nano-FTIR spectra (Fig. 2f) were acquired in 0.1 mg/ml SF solutions, which, as shown above, lead to multilayer growth. The results show that the intensity of the peak in amide II region (1480-1600 cm$^{-1}$) starts to increase dramatically after 50 minutes of incubation and reaches its maximum after 70 minutes. In contrast, only a weak signal from the water bending mode appears in amide I region (1600-1700 cm$^{-1}$)[22,23], where ATR-FTIR data collected on a bulk silk sample records a strong peak (Fig. 2g). This effect is exclusively due to the polarization of the near-field IR beam, which is exclusively responsive to vibrational modes characterized by transition dipole moments perpendicular to the surface[22,24]. The low ratio of the amide I band intensity to that of the amide II band indicates a preferential vibration of the C=O bond parallel to the surface (Fig. 2g), which is consistent with the C=O bond direction seen in the MD simulations (Extended Data Fig. 5d) and implies a uniform configuration of the SF molecules within the lamellae. Taken together, these two features in the nano-IR spectra imply that the silk molecules themselves are well-ordered within the lamellae.

The evolution in peak intensity of the amide bands over time reflects the structural rearrangement of the SF molecules during lamella formation. The initial increase in the amide II region indicates rapid development of the β-sheet conformation. Conversely, the amide I region displays no changes in IR signal intensity, indicating that there are no unfolded proteins on the surface, even in the early stages of lamella formation. These observations imply that the SF proteins add to the lamellae directly from solution and adopt the β-sheet structure at the time of attachment, leading to significant longitudinal growth (Extended Data Fig. 6). Thus, providing compelling evidence that the assembly of the lamellae follows the conventional formation pathway of 3D silk fibers grown in bulk solution.

**Two-step formation of silk multi-layers**

When the SF concentration is further increased to ≥ 0.2 mg/ml, a new, unstructured layer begins to adsorb onto the underlying lamellar layer (Fig. 3a). This adsorption is accompanied by the appearance of the amide I signal at 1640-1655 cm$^{-1}$ implying the presence on the graphene surface of unfolded SF proteins, which are known to be mainly random coil in their unfolded state, (Extended Data Fig. 5e) and can thus be detected by nano-FTIR through their C=O stretching vibration. Domains of the unstructured film coexist with lamellar domains and can be distinguished both by their topography (Fig. 3b) and their phase (Fig. 3c), with the latter generally corresponding to mechanical stiffness[25]. The lamellae in these layers still lie along the three armchair directions of the underlying HOPG and are higher by ~ 0.1 nm than the unstructured film (Fig. 3d, e) with which they form a sharp boundary (Fig. 3f and Extended Data Fig. 7a-d). Time series of in-situ AFM images show that the lamellar domains and the sharp boundaries are a consequence of a phase transition from the unstructured film to lamellae, which is accompanied by the increase in height (Fig. 3g-i, and Extended Data Fig. 7e). Hence, the unstructured film is a metastable phase consisting of unfolded SF molecules, which converts into the structured film through folding and reorganization into β-sheet lamellae. This defines the second pathway of lamellar SF multilayer assembly (Fig. 3j). Moreover, these observations follow the generally process of silk formation through a structural transition from random coil to β-sheet conformation[26,27].

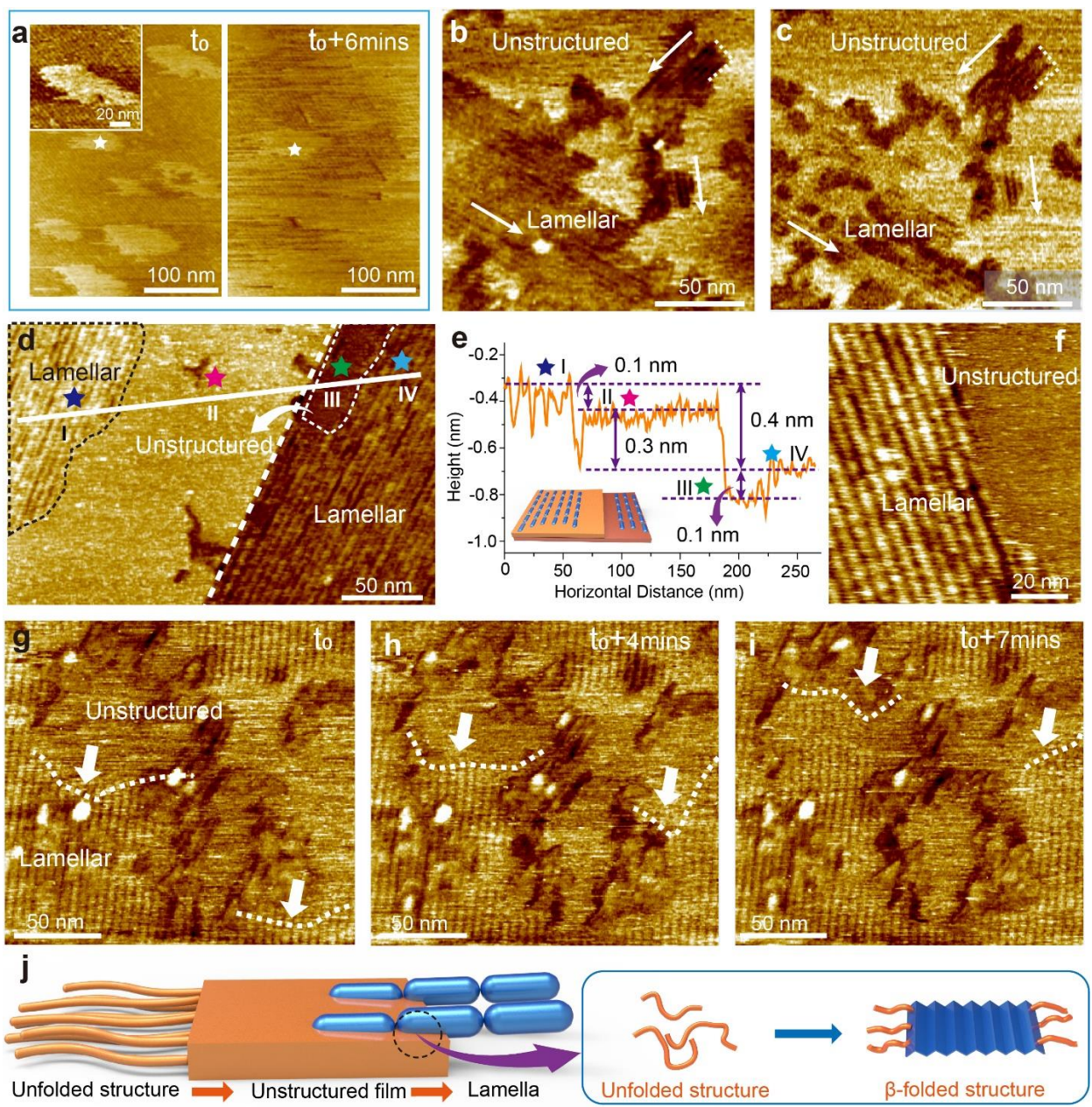

**Fig. 3 | The two-step growth process of lamellar SF films. a,** Consecutive AFM height images of an unstructured film forming from 0.2 μg/ml SF solution. Inset shows that the unstructured film is deposited on a lamellar layer. **b, c,** AFM height and phase images during SF assembly from 1 μg/ml SF solution. White arrows indicate the three directions of the lamellae. Both lamellar and unstructured domains can be easily identified in both, indicating that the domains differ in stiffness. **d,** AFM height image showing two complete layers in which lamellar and unstructured regions co-exit. The white dashed line delineates the edge of the upper layer. The black dashed boundary denotes a lamellar region in the top layer, while the white dashed boundary denotes an unstructured region in the bottom layer. **e,** Height profile along the solid white line in (**d**). Blue, pink, green, and cyan stars denote different structural regions: lamellar (blue and cyan) and unstructured (pink and green). The inset illustrates the model of multilayer SF films with ordered and disordered regions. **f,** High-magnification AFM height image of the boundary between lamellar and unstructured domains. **g-i,** Consecutive AFM images of assembly from 1 μg/ml SF solution showing the phase transition from unstructured to lamellar. The white dashed lines mark the boundaries between the two. **j,** Proposed model for the two-step process of the lamellar SF film formation.

To gain insights into the structural dynamics that accompany the unstructured-to-lamellar transition, we used *in situ* nano-FTIR to observe the conformation state of the SF. The IR signal in the amide I region was detected, but cycled back and forth between the peaks at ~ 1630 cm$^{-1}$ and ~ 1650 cm$^{-1}$ associated with the β-sheet and unfolded SF conformation, respectively (Fig. 4a, b), and did so on a timescale (5-15 min) similar to that seen in AFM during the transition from the unstructured to lamellar film. This cyclical process facilitates the accumulation of well-ordered SF lamellar layers in the z-direction even when the SF concentration is increased by 6x (to 3 μg/ml). (Extended Data Fig. 7f, g). The intensity of the amide I peak centered at 1630 cm$^{-1}$ is then higher than that of the amide II bands (Extended Data Fig. 8), demonstrating the attainment of a thick 2D SF lamellar film (Extended Data Fig. 7f, g). Thus, these data record the cyclical process of adsorption of an unstructured film followed by transformation to a lamellar film (schematic in Fig. 4a) and show that, indeed, the unstructured film consists of unfolded SF proteins and the transition to the lamellar film is accompanied by folding into the β-sheet conformation.

**Discussion**

The above findings show that, in contrast to bulk SF fibrils, which are both mesoscopically and macroscopically disordered (Fig. 4c), SF self-assembles in an epitaxial manner into well-ordered 2D lamellar films on HOPG, with the building units of the film exhibiting the same secondary structure as the β-sheet crystallites in the bulk form. While two assembly pathways arise due to the competition between the adsorption rate of unfolded SF molecules and the folding and assembly into the lamellae, the outcome remains a well-ordered 2D lamellar film (Fig. 4d-j).

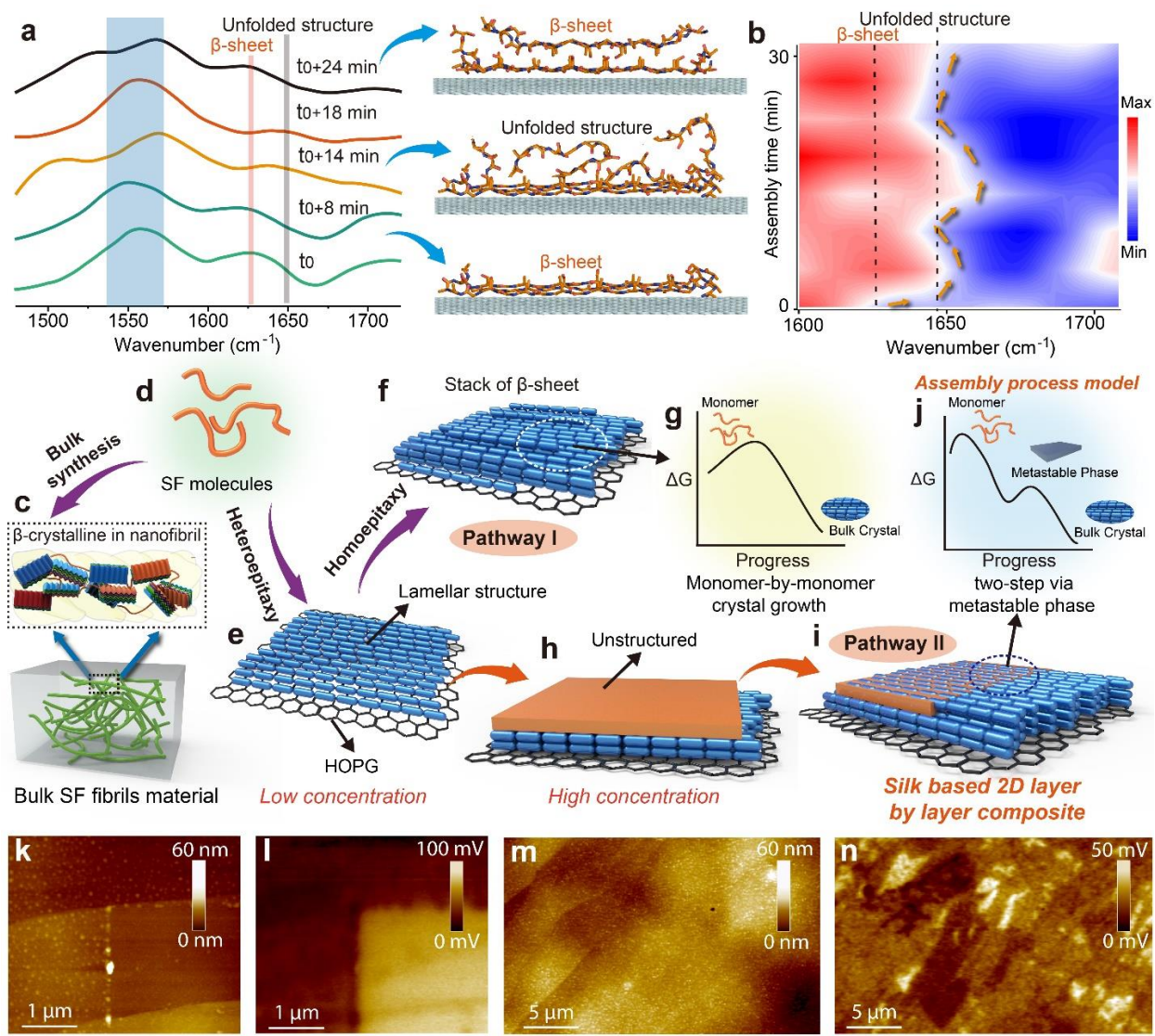

**Fig. 4 | The mechanism of 2D SF assembly at the water-HOPG interface**. **a**, *In situ* nano-FTIR spectra showing the change in secondary structure during assembly on graphene from high-concentration SF solution (0.5 µg/ml), along with an illustration of the process. **b**, Corresponding color map intensity of the nano-FTIR spectra as a function of time. **c**, Depiction of bulk SF fibrils structure at mesoscopic and macroscopic scales. **d-j**, Proposed scheme of 2D SF assembly at the water-HOPG interface. **k**, **l**, AFM

height image (**k**) and corresponding SKPM image (**l**) showing the top SF film was selectively removed from the bright region in (**l**) through scraping to reveal the underlying HOPG where the surface potential is significantly higher. **m**, **n**, AFM height image (**m**) and corresponding SKPM image (**n**) of SF film on HOPG.

Given the high degree of crystalline packing in these films, in analogy to films of conjugated polymers, we might expect them to exhibit strong orbital overlap[28], which has been shown to improve the electronic performance of film-based devices, such as perovskite solar cells[29]. Importantly, heteroepitaxial assembly of 2D SF films also occurs on other 2D VdW surfaces, such as $MoS_2$, (Extended Data Fig. 9a-c) (though not on the hydrophilic surface of mica, Extended Data Fig. 9d-f). Consequently, this 2D phase of SF may be a useful construct for modulating the electronic properties of vdWs materials in general. Indeed, surface potential measurements using scanning Kelvin probe microscopy (SKPM) on HOPG demonstrate a 60 mV decrease in potential following SF assembly (Fig. 4k, l). Such a decrease has been shown to significantly reduce the barrier to metal ion migration and increase the transport rate of charged particles in protein films[8]. In addition, because differences in SF film thickness lead to variations in surface potential (Fig. 4m, n) systematic design of multi-layers can enable programmable modulation of electronic properties. Moreover, although domains with orientations distinct from the dominant armchair direction are unstable in the presence of the latter and are eliminated through coarsening (Extended Data Fig. 10), the fact that such domains still form suggests sufficiently weak film-substrate binding that a fully formed, crosslinked film could be twisted relative to the underlying vdW substrate to achieve additional modulation of the electronic properties, just as interlayer twisting of the vdW materials has been used for this purpose[30,31]. Taken together, these results and considerations suggest that the ability to create highly-ordered 2D SF layers on multiple vdW materials provides a novel strategy for both extending and improving the performance of silk in electronic and optical applications.

## Methods

### Material preparation

Silk fibroin (SF) was regenerated from the cocoons of domesticated B. mori silkworms (Guangxi Sericulture Technology Co., Ltd.) as previously described[8]. The fresh silk fibroin solution was injected into liquid nitrogen through a syringe pump to form small globules and then freeze-dried for 3 d. The dried SF globules were stored at -20 °C before further usage. Sodium bicarbonate (NaHCO3) and isopropanol were obtained from Sinopharm Chemical Reagent. Lithium bromide (LiBr) were provided by Aladdin Industrial Corporation. HOPG (HOPG grade ZYB, 0.8° 10 x 10 mm x 1mm) was purchased from Ted Pella, Inc.

### In-situ AFM Characterization.

SF powders were mixed with 10 mL nuclease-free water (Ambion, USA), was purified with 0.2 μm filter to remove aggregate and insoluble SF. 80 μL of SF solution was added on top of a freshly cleaved HOPG surface at room temperature. In-situ images were captured using silicon probes (SNL, k: 0.12 N/m, tip radius: 2 nm; Bruker) and silicon nitride probes (SCANASYST-FLUID, k: 0.7 N/m, tip radius: 20 nm; Bruker) under tapping mode with a Cypher ES AFM (Asylum Research). Images were analyzed using Scanning Probe Image Processor (SPIP) (Image Metrology, Denmark) and Gwyddion SPM data analysis software.

### In-situ Nano-FTIR Characterization.

The present study employed the technique of in-situ Nano-FTIR characterization, building upon previous research[22]. The measurements using Nano-FTIR were conducted at beamline 2.4 of the Advanced Light Source, located at the Lawrence Berkeley National Laboratory. During the experimental setup, infrared light was carefully directed onto the Pt-coated AFM tip within a neaSNOM (Near-field Scanning Optical Microscopy) system. Detailed information regarding the fabrication of the graphene membrane can be found in prior publications. For the experimental procedure, the SF solution was enclosed within the graphene liquid cell immediately prior to the commencement of the experiment, denoted as time 0. To establish reference spectra, samples with flat spectral responses were utilized, including either Au-coated Si or graphene on the Au-coated SiN membrane. Tapping-mode operation was employed, operating at the fundamental resonance frequency of the cantilever (ranging from 250 to 350 kHz) with a free oscillation amplitude varying between 70 and 90 nm. The amplitude setpoint was maintained at approximately 80%. To eliminate the far field nonlocal scattered background, the scattered nearfield signal was extracted using a lock-in amplifier that was tuned to the second and higher harmonics of the cantilever oscillation.

**In-situ Circular dichroism spectrum**

The circular dichroism (CD) spectrum of SF (0.2 mg/mL) was collected on a Circular Dichroism Spectropolarimeter (Applied Photophysics Limited) at room temperature lasted 80 h.

**Attenuated Total Reflection Fourier Transform Infrared (ATR-FTIR)**

The SF solution (0.27 mg/mL) was casted on HOPG and incubated for 1 h, then remove the water and dried under fume hood. The ATR-FTIR measurements were conducted by a FTIR Perkin Elmer Frontier with a diamond crystal ATR module. The spectra were collected with 4 cm-1 resolution in the spectral range from 4000 cm-1 to 400 cm-1 averaging over 64 scans and normalized to the light intensity.

**Scanning Kelvin Probe Microscopy (SKPM)**

Kelvin probe force microscopy was carried out using CypherS AFM (Asylum Research, CA) in the air with Platinum coated AFM cantilevers (Multi75E, Budget Sensors). In SKPM measurements, the sample was grounded. The offline data process was done with SPIP software (Image Metrology, Denmark)

**Molecular Dynamics Simulations**

Molecular dynamics simulations were run with the GROMACS 2021.34 engine and PLUMED 2.7 plugin, using the CHARMM36 and Interface force fields as well as the TIP3P water model. The Parrinello-Rahman barostat was used for simulations in the NPT ensemble, and temperature was coupled using a Noose-Hoover extended ensemble. All systems were energy minimized before equilibration in the NVT and NPT ensembles. Production simulations were run in the NVT ensemble. Folded SF protein structures were generated with PyMOL, and HOPG structures were generated with CHARMMGUI. To replicate experimental conditions, all graphite surfaces were composed of five layers. Analysis was performed with the NumPy, MDAnalysis, and MDTraj Python libraries, and figures were generated with Matplotlib, PyMOL, and VMD. Simulations were run on 2080ti GPUs on the University of Washington Klone cluster. All input files and scripts necessary to replicate the simulations and analysis performed in this study are available in the Supporting information. The distribution of angles between C=O bonds and graphite was calculated by defining vectors along C=O bonds and calculating the angle between these vectors and the Z-normal vector with the following equation:

$$\theta = \cos^{-1} \frac{a \cdot b}{|a||b|}$$

Where a is the vector formed by C=O bonds and b is the Z-normal vector. Because the Z-normal vector is perpendicular to the XY plane, the resulting angles can easily be converted to angles between C=O bonds and the XY plane by subtracting 90o. The XY plane was chosen as a representation of graphite because the surface is perfectly flat in the XY plane during simulation.

**Acknowledgments:** This project was supported by the US Department of Energy (DOE), Office of Science (SC), Office of Basic Energy Sciences (BES), Division of Materials Science and Engineering, Biomolecular Materials Program at Pacific Northwest National Laboratory (PNNL) under award FWP 65357. PNNL is a multiprogram national laboratory operated for DOE by Battelle under contract number DE-AC05–76RL01830. Molecular dynamics simulations were supported by the DOE BES Energy Frontiers Research Centers program through CSSAS: The Center for the Sciences of Synthesis Across Scales, at the University of Washington under award number DE-SC0019288. High resolution AFM imaging was performed in the Molecular Analysis Facility (MAF), University of Washington (UW). The MAF is part of the National Nanotechnology Coordinated Infrastructure (NNCI), a National Science Foundation-funded effort to coordinate nanoscale research and development activities across the United States and is supported by NNCI-2025489 and NNCI-1542101. SKPM research was conducted as part of a user project (CNMS2022-B01603) at the Center for Nanophase Materials Sciences (CNMS), which is a US DOE, SC BES User Facility at Oak Ridge National Laboratory. Nano-FTIR was performed at beamline 2.4 and 5.4 of the Advanced Light Source (ALS), a DOE SC BES User Facility operated by Lawrence Berkeley National Laboratory under contract number DE-AC02-05-CH11231).

**Author contributions**: C.S., S.Z., J.D.Y. and X-Y.L. designed the project. C.S. prepared the samples and carried out the AFM experiments, with the assistance from S.Z., and executed the FTIR experiments with the assistance of X.Z. and M.B.S. M.Z. performed the MD simulations with the assistance of J.P. S.Z. collected the SKPM data. C.S., M.Z., S.Z., J.P. and J.D.Y. wrote the manuscript and all authors discussed the results and commented on the manuscript.

**Competing interests:** Authors declare that they have no competing interests.

**Data and materials availability**: All data are available in the main text or the supplementary materials.

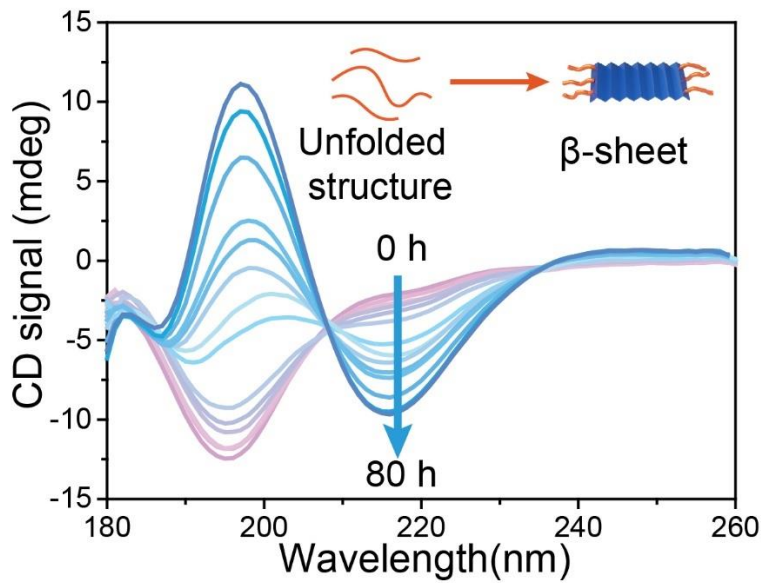

**Extended Data Fig. 1 | Circular dichroism (CD) spectrum of SF solution**. The observation of a negative peak in the region around 190–200 nm indicates that the freshly prepared SF solution primarily consists of unfolded structures, predominantly in the form of random coils. Over time, these unfolded structures undergo a gradual transformation into β-sheets, which represents a natural phenomenon in the nucleation and growth process of silk materials. Note: β-sheets shows a negative band at 218 nm and a positive one at 196 nm.

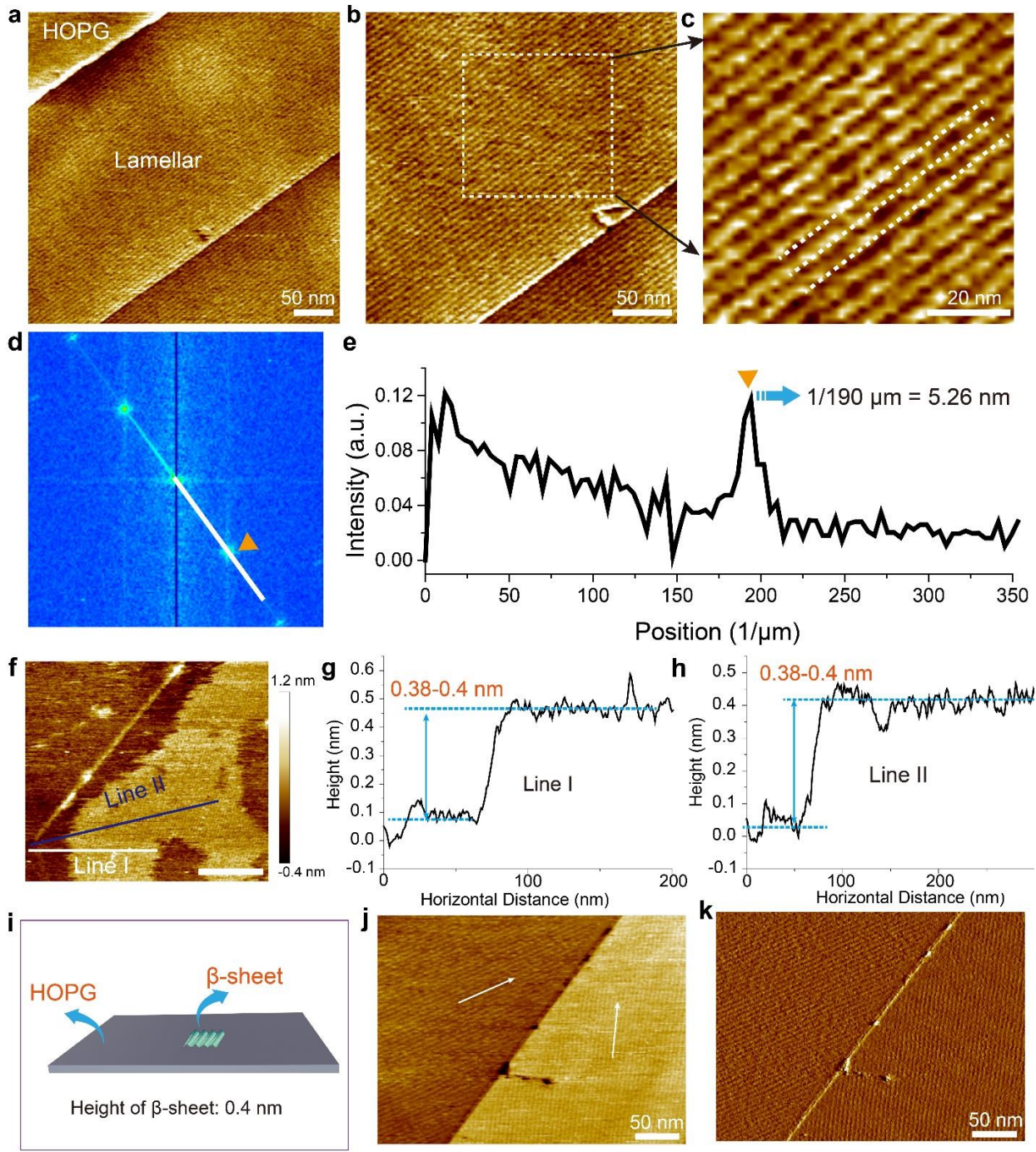

**Extended Data Fig. 2 | The structure information of SF lamella. a-e,** Periodic width of lamellar structure of SF. Topography map of different scale (**a-c**), FFT image (**d**) and 1D Fourier profile (**e**). **f-i,** Height information of lamellar structure of SF. Topography map (**f**), scale bar is 100 nm. Height profile for line 1 (**g**). Height profile for line 2 (**h**). Schematic drawing of β-sheets structure on HOPG (**i**). **j, k,** The orientation of SF lamellar structure. SF assembled into lamellae which formed at an angle of 120° to each other, indicating

aligned along three equivalent directions on HOPG. Height image (**j**). Amplitude image (**k**).

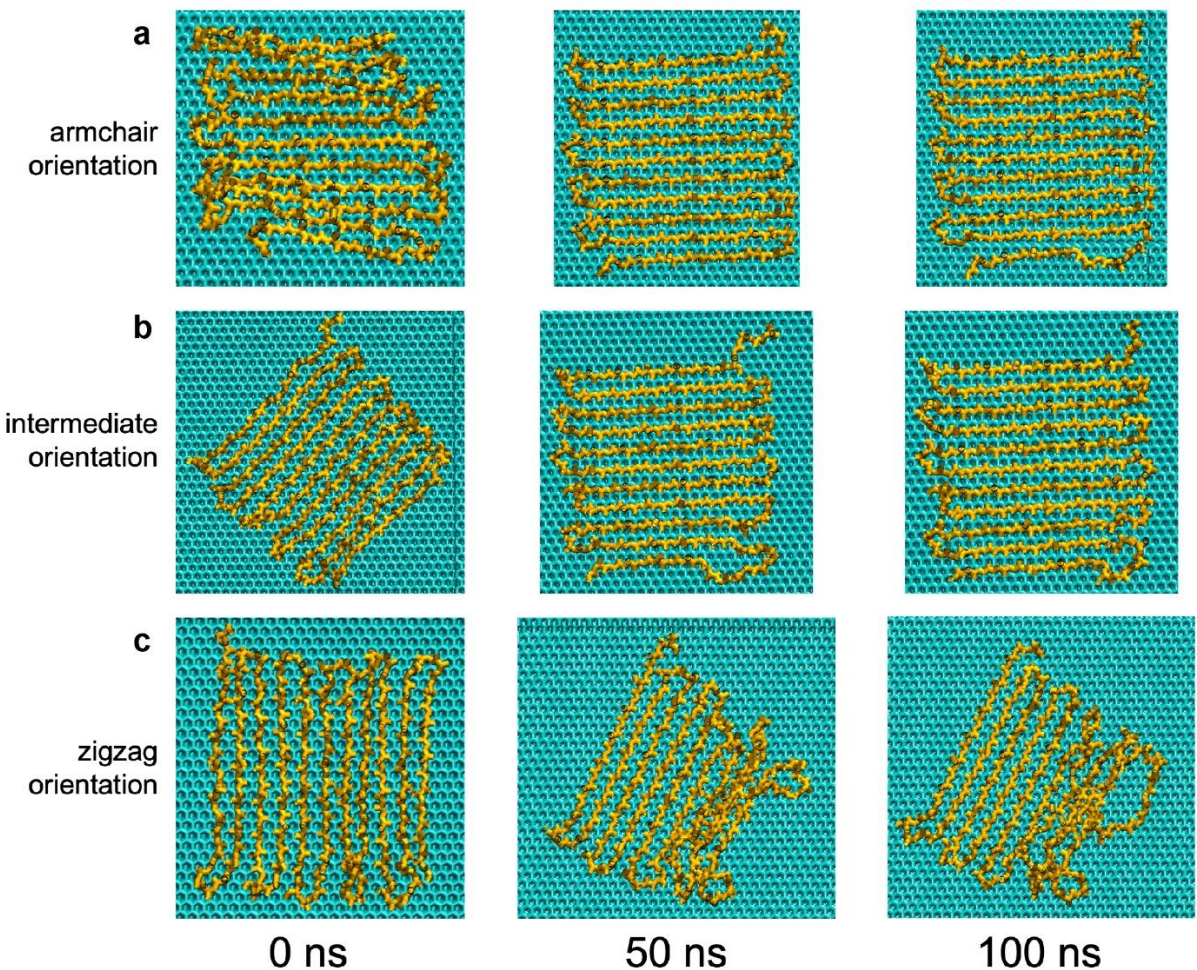

**Extended Data Fig. 3 | Simulations of SF β-sheets binding to HOPG**. SF β-sheet was placed ~1 nm above graphite and allowed to bind. The protein initially aligned along the armchair edge. (**a**) did not rotate. However, the proteins aligned to the intermediate and zigzag orientations, (**b** and **c**) rotated to align to the armchair edge during binding, and remained at this orientation throughout the simulation.

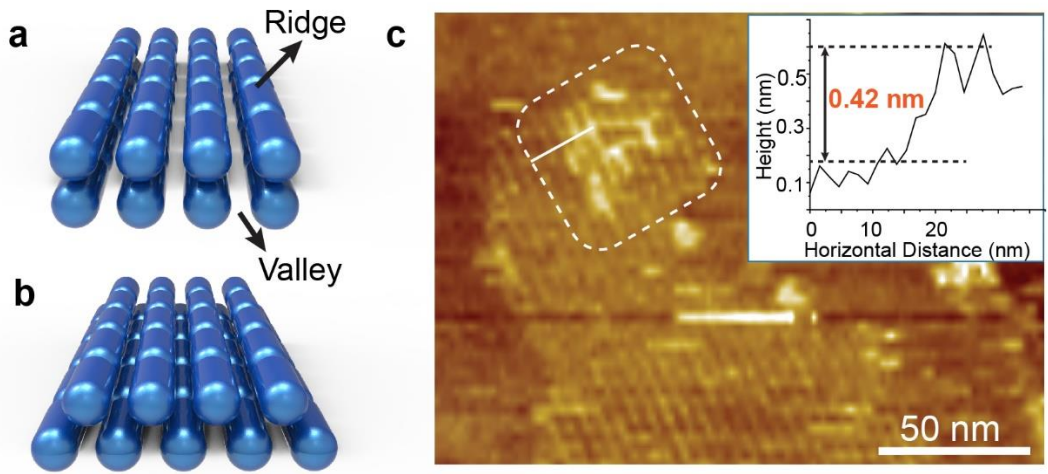

**Extended Data Fig. 4 | The co-align growth of lamellar structure**. **a**, **b**, The lamella model of co-align growth, the upper lamella grew along the "ridge" of the underneath lamellar structure (**a**) rather than lateral dislocation to the "valley" of the underneath lamellar structure (**b**), indicating SF β-strands precisely matching in vertical orientation induced by intermolecular interaction inside groups. **c**, A high-magnification AFM image of the multilayer lamella structure. The white dashed circle highlights the detail of the stacked parallel lamella. Inset is the height line-profile along the white line in the white dashed circle.

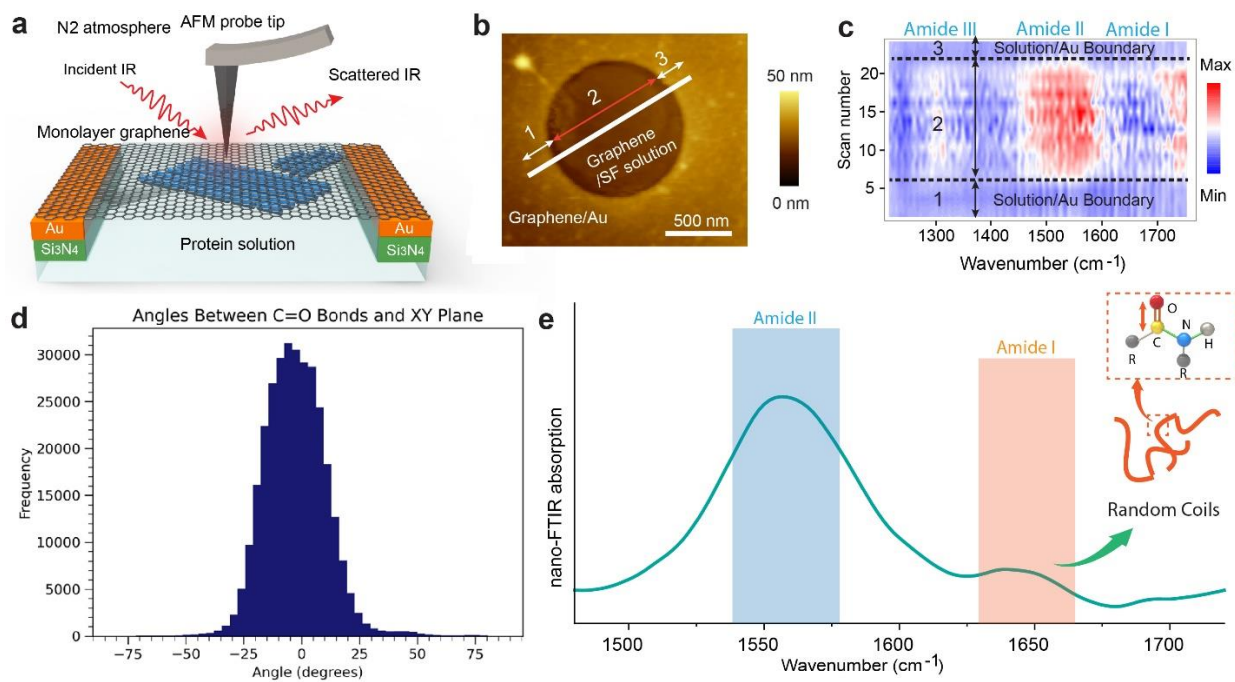

**Extended Data Fig. 5 | The nano-FTIR analysis of lamellar structure**. **a**, Schematic drawing of the nano-FTIR experiment. The monolayer graphene separates the air and protein solution. **b**, **c**, Spatially resolved chemical mapping of SF assemblies. (**b**) Images of the total scattered optical amplitude of suspended graphene and surrounding area in the cell filled with SF (0.1 μg/mL). (**c**) Corresponding color map representations of the nano-FTIR spectral intensities of the amide I, II, and III mode regions of the SF. The nano-FTIR profiles were acquired at positions along the white line in image (**b**). The number 1,2 and 3 represent the location of graphene/Au, graphene/SF solution and graphene/Au, respectively. **d**, Distribution of angles between vector defined by protein backbone C=O bonds and the XY plane. The XY plane was chosen as a proxy for graphite as the surface is exactly flat along the XY plane in the simulation. Angles were calculated only for non-turn residues, as turns are lifted and therefore preclude C=O bonds from lying flat. The distribution spread is likely a result of thermal fluctuations in atomic positions. **e**, Nano-FTIR spectra of an assembled SF structure in 0.5 μg/mL SF solution. The detection of the amide I signal during assembly above a concentration of 0.1 μg/mL SF solution serves as evidence for the presence of an unfolded structure (such as random coil) on the surface.

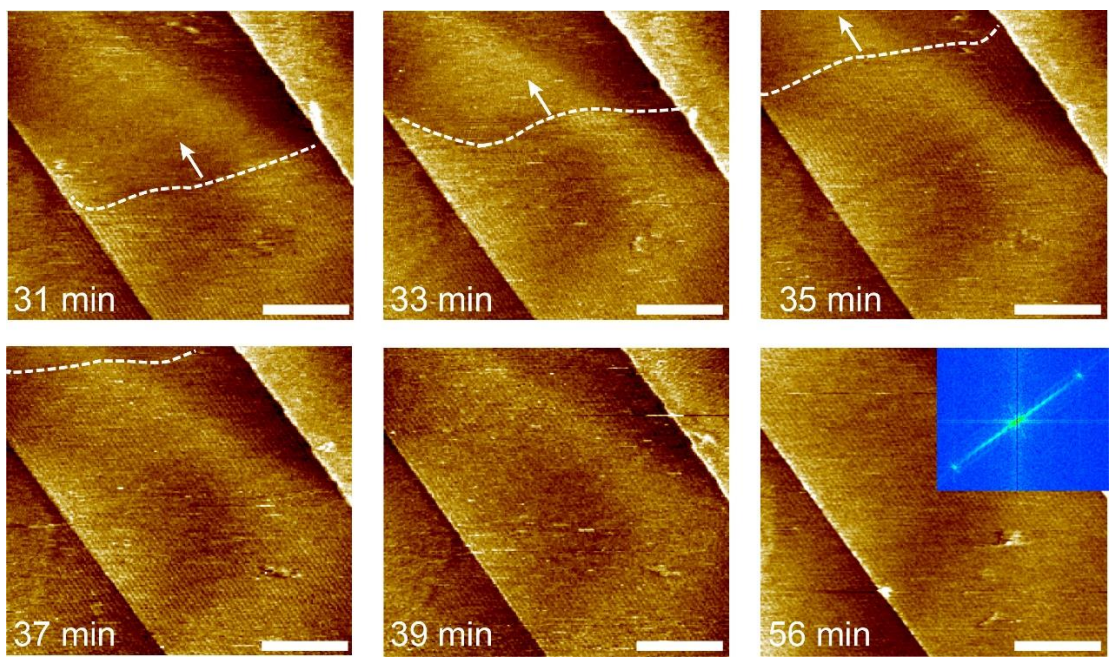

**Extended Data Fig. 6 | AFM images of SF self-assembly at different growth times.**
The illustration in 56 min image is the corresponding FFT image, scale bar is 100 nm.

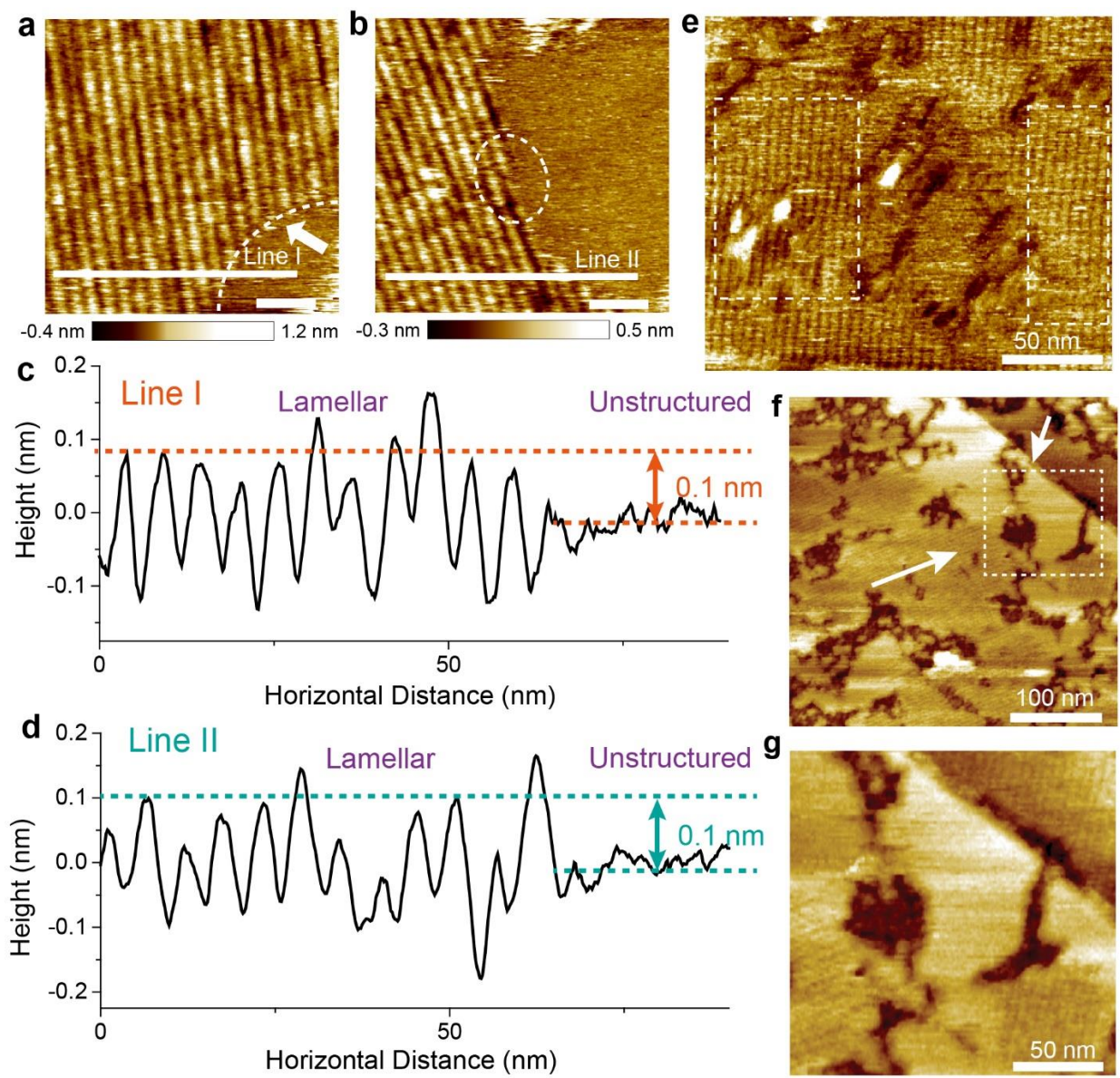

**Extended Data Fig. 7 | SF lamellae grown at high concentration. a-d,**The height information of two structures. (**a**, **b**) Topography map two structures. **c**, The height data of two structures (image **a**, line I). **d**, The height data of two structures (image **b**, line II). Scale bar is 20 nm. **e**, The AFM image of SF assembly. The white dash box in AFM image shows all the unstructured had been transformed into lamellar structure. **f**, **g**, The self-assembled data of SF with high concentration. The concentration of SF is 3 μg/mL. (**f**) The topography map of SF after 59 min of self-assembly, the inset is FFT image. (**g**) The corresponding topography map of white box in image (**f**).

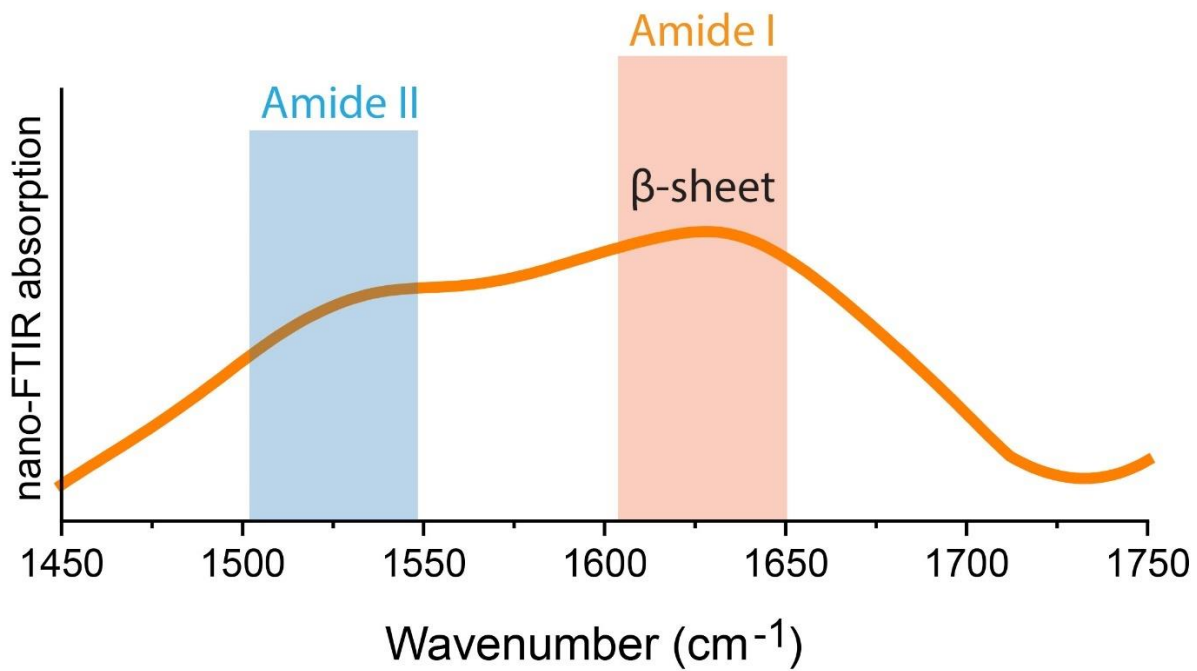

**Extended Data Fig. 8 | Nano-FTIR spectra of an assembled SF structure in 3 μg/mL SF solution.** The intensity of amide I signal higher than amide II during assembly in a concentration of 3 μg/mL SF solution, and the peak of amide I centered at 1630 cm$^{-1}$, representing the formation of more β-sheet structure, that is a thicker lamellar layer.

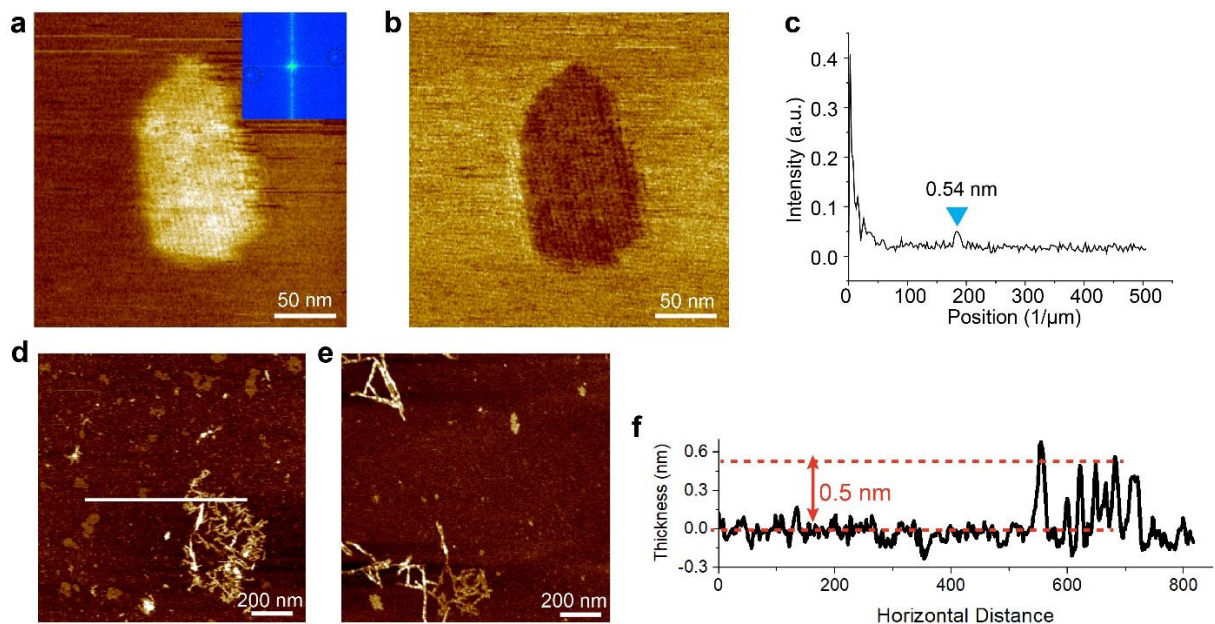

**Extended Data Fig. 9 | SF assemble on different substrate. a-c**, The lamellar structure on MoS2. (**a**) Morphology image. (**b**) Phase image. (**c**) 1D Fourier profile. The periodicity of lamellar on MoS2 is same with HOPG, indicating a similar assemble process perform at the 2D surface. **d-f**, There is no lamellar structure on Mica. The morphology image (**d**, **e**). The height of SF structure on Mica is around 0.5 nm which had defined as SF monomer (**f**).

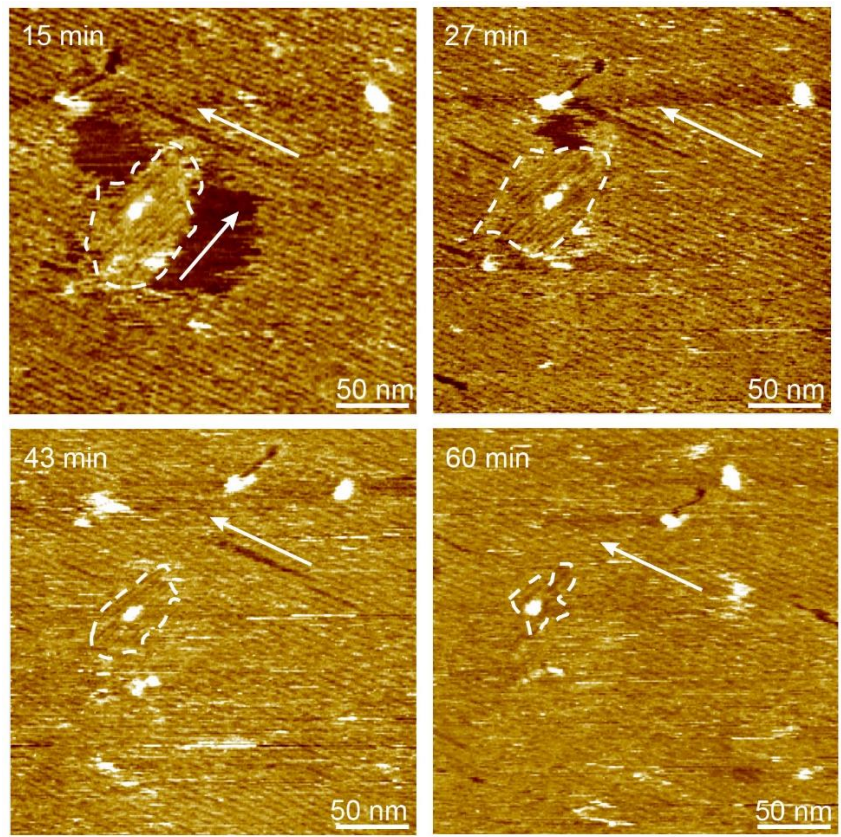

**Extended Data Fig. 10 | The lamellae twist the orientation.** In situ AFM phase images showing lamellae twist the orientation as nearby lamellae grew. The growth direction in white dash box gradually replaced by another direction.

# Supplementary Materials for

## Two-dimensional silk

Chenyang Shi[1,2], Marlo Zorman[3], Xiao Zhao[4,5], Miquel B. Salmeron[4], Jim Pfaendtner[6], Xiang Yang Liu[7]*, Shuai Zhang[1,2]*, James De Yoreo[1,2]*

Corresponding author: james.deyoreo@pnnl.gov (J.J.D.Y.); zhangs71@uw.edu (S.Z.); liuxy@xmu.edu.cn (X.L.).

## Supplementary Text

We tested orientation preferences of SF β-sheet on HOPG as a validation of the force fields (Extended Data Fig. 5d). SF was placed ~1 nm above the surface, and the backbone was restrained using the PLUMED ANTIBETARMSD collected variable. Three initial orientations were tested: one in which the SF β-sheet was aligned with the armchair edge, one in which it was aligned with the zigzag edge, and an intermediate orientation. The SF β-sheet bound to the surface completely within 5 ns in all three trials. The SF β-sheet displayed a strong preference for the armchair orientation. Both the intermediate and zigzag orientations aligned to the armchair edge during binding and stayed at this orientation for the remainder of the 100 ns simulation. The armchair orientation did not rotate during binding. Note that these simulations do not provide any mechanistic information and are simply a validation that the force field is able to reproduce orientational preference.

| System | Height (Å) |
|---|---|
| Face-to-face double layer | 3.94 ± 0.077 |
| Face-to-back double layer | 3.26 ± 0.028 |

**Table S1. The height of a lamellae layer was determined by extracting the z-coordinate of the layer's non-turn residue backbone center of mass.** For the monolayer lamellae, thickness was calculated relative to the z-coordinate position of carbon atoms in the first layer of graphene. For the bilayer lamellae, thickness was calculated by subtracting the height of the lower layer from that of the upper layer. Thickness was averaged over the last 90 ns of production simulation time for all systems.

| System | Notes | Size (nm) | Simulation Time | Replicas |
|---|---|---|---|---|
| Single protein alanine SCs up along armchair | Protein secondary structure restrained during equilibration/binding | 10.4x10.3 x10.5 | 125 ps NVT, 50 ps NPT, 100 ns NVT | 3 |
| Single protein alanine SCs up along zigzag | Protein secondary structure restrained during equilibration/binding | 10.4x10.3 x10.6 | 125 ps NVT, 50 ps NPT, 100 ns NVT | 1, 2 IP |
| Single protein alanine SCs up along intermediate angle | Protein secondary structure restrained during equilibration/binding | 10.4x10.3 x10.7 | 125 ps NVT, 50 ps NPT, 100 ns NVT | 1, 2 IP |
| Single protein alanine SCs down along armchair | Protein secondary structure restrained during equilibration/binding | 10.4x10.3 x10.8 | 125 ps NVT, 50 ps NPT, 100 ns NVT | 3 |
| 1x4 fiber alanine SCs up | Proteins interacted across y-axis | 9.9x20.9x 10.7 | 125 ps NVT, 50 ps NPT, 100 ns NVT | 1, 2 IP |
| 1x4 fiber alanine SCs down | Proteins interacted across y-axis | 9.9x20.9x 10.7 | 125 ps NVT, 50 ps NPT, 100 ns NVT | 1, 2 IP |
| 2x2 bilayer upper layer alanine SCs up | Inter-protein waters placed during solvation were deleted | 11.3x10.7 x6.7 | 125 ps NVT, 300 ps NPT, 100 ns NVT | 1, 2 IP |
| 2x2 bilayer upper layer alanine SCs down | Inter-protein waters placed during solvation were deleted | 11.3x10.7 x6.7 | 125 ps NVT, 300 ps NPT, 100 ns NVT | 1, 2 IP |
| Unfolded protein on lamallae | System size decreased after binding | 9.9x20.5x 8.7, 8.4x10.7x 6.5 | 125 ps NVT, 50 ps NPT, 1000 ns NVT | 1, 2 IP |
| Unfolded protein on hopg | None | 10.4x10.3 x7.5 | 125 ps NVT, 50 ps NPT, 1000 ns NVT | 1, 2 IP |

**Table S2. The basic parameter of the simulations model.**